\documentclass[letterpaper]{article}
\usepackage{aaai}
\usepackage{graphicx}
\usepackage{fixbib}
\usepackage{times}
\usepackage{helvet}
\usepackage{courier}
\usepackage{epsfig}
\usepackage{amssymb} 
\usepackage{amsmath}
\usepackage{amsfonts}
\usepackage{algorithm}
\usepackage{algorithmic}
\usepackage{mdwlist}
\usepackage{multirow}
\usepackage{color}
\usepackage{subfigure}
\usepackage{url}

\newenvironment{definition}{\begin{trivlist}
\item[]}{\end{trivlist}}

\begin{document}
%
\title{Learning User-specific Latent Influence and\\ Susceptibility from
Information Cascades} 
\author{Yongqing Wang$^{*}$, Huawei Shen$^{\dag}$, Shenghua Liu$^{\dag}$
and Xueqi Cheng$^{\dag}$\\
CAS Key Laboratory of Network Data Science and Technology, \\ Institute of
Computing Technology, Chinese Academy of Sciences, Beijing 100190, China\\
$^*$wangyongqing@software.ict.ac.cn,
$^\dag$\{shenhuawei,liushenghua,cxq\}@ict.ac.cn\\
}

\maketitle
\begin{abstract}
Predicting cascade dynamics has important implications for understanding
information propagation and launching viral marketing. Previous works mainly
adopt a pair-wise manner, modeling the propagation probability between pairs of
users using $n^2$ independent parameters for $n$ users. Consequently, these
models suffer from severe overfitting problem, especially for pairs of users
without direct interactions, limiting their prediction accuracy. Here we
propose to model the cascade dynamics by learning two low-dimensional
user-specific vectors from observed cascades, capturing their influence and
susceptibility respectively. This model requires much less parameters and thus
could combat overfitting problem. Moreover, this model could naturally model
context-dependent factors like cumulative effect in information propagation.
Extensive experiments on synthetic dataset and a large-scale microblogging
dataset demonstrate that this model outperforms the existing pair-wise models
at predicting cascade dynamics, cascade size, and ``who will be retweeted''.
\end{abstract}

\section{Introduction}
Social media is revolutionizing the dissemination of information via its great
ease in information delivery, accessing and filtering. In social media, users
could post original messages or forward messages that they see from other
users. Information propagation proceeds along social relationships between
users, explicit or implicit, forming cascade dynamics. Modeling and predicting
the cascade dynamics has important implications to understanding
information propagation and launching viral marketing in social media. The key
for this problem is inferring the interpersonal influence between users or
estimating the probability that information propagates between them,
fundamental to influence
maximization~\cite{kempe2003maximizing,ChenKDD2009,ChengCIKM2013}, social
recommendation~\cite{huang2012exploring,ma2009learning}, and viral
marketing~\cite{RichardsonKDD2002,leskovec2007dynamics}.

Existing studies mainly aim to determine the propagation probability of
information between all pairs of users, based on structure of social network,
the record of information cascade, and demographic/content characteristics of
users. Kempe et al.~\cite{kempe2003maximizing} implemented the
independent cascade model for information propagation, assuming a uniform propagation
probability or a degree-modulated propagation probability. Goyal et
al.~\cite{goyal2010learning} provided two static models in terms of Bernoulli
distribution and Jaccard index, and learned temporal factors to maximize
likelihood of cascades. Saito et al.~\cite{saito08} learned the propagation
probability for independent cascade model in terms of expectation maximization
of cascades. Artzi et al.~\cite{Artzi2012} estimated the propagation probability
by exploiting demographic and content characteristics. These methods all adopt
a pair-wise manner, modeling the propagation probability between pairs of users
using $n^2$ independent parameters for $n$ users. Consequently, these models
suffer from severe overfitting problem, limiting their prediction accuracy. For
example, for a pair of users without observed interactions, these methods take
the propagation probability between them as zero, indicating that it is never
happened to propagate information between the two users in the future. To the
best of our knowledge, we lack a model that could concisely model the
interpersonal influence and accurately predict the cascade dynamics in
large-scale social networks.

In this paper, we propose to model the cascade dynamics by learning two
low-dimensional latent vectors for each user from observed cascades, capturing
her influence and susceptibility respectively. In this latent influence and
susceptibility (LIS) model, the propagation probability that one user forwards
a piece of information is determined by the product of her activated neighbors'
influence vectors and her own susceptibility vector. The benefits of this model
are three-fold: (1) It directly models user-specific influence and
susceptibility, instead of the interpersonal influence between all pairs of
users. Thus it requires much less parameters, and pairs of users are no longer
independent, effectively combating the overfitting problem for pairs of users
without direct interactions; (2) It could naturally capture context-dependent
factors like cumulative effect in information propagation. For a target user,
context means her previous exposures to the same message.
In this model, one user's previous exposures to a message improve the
probability that she propagates the message, flexibly combining the benefits of
both cascade model and threshold model; (3) It is applicable to scenarios with
explicit or implicit social networks, since it learns user-specific influence
and susceptibility from the observed information cascades rather than the
underlying social networks like many existing methods.

We evaluate the proposed LIS model by extensive experiments on synthetic
dataset and a large-scale microblogging dataset from Sina Weibo, the largest
social media in China. Compared with several widely-used methods that work in pair-wise manner,
LIS model consistently outperforms them at predicting the dynamics of
cascades. Moreover, the learned user-specific influence and susceptibility
vectors provide us a quantitative way to understand topic-related interpersonal
influence in information propagation.

\section{Problem formulation}
\label{sec:problem_formualtion}

In this paper, we focus on the problem of inferring users' influence and
susceptibility from detailed records of message cascades. Before diving into the
details of the proposed model, we first clarify the two main motivations
underlying our model.

\begin{figure}[t]
\centering
\subfigure[]{\includegraphics[width=0.19\textwidth]{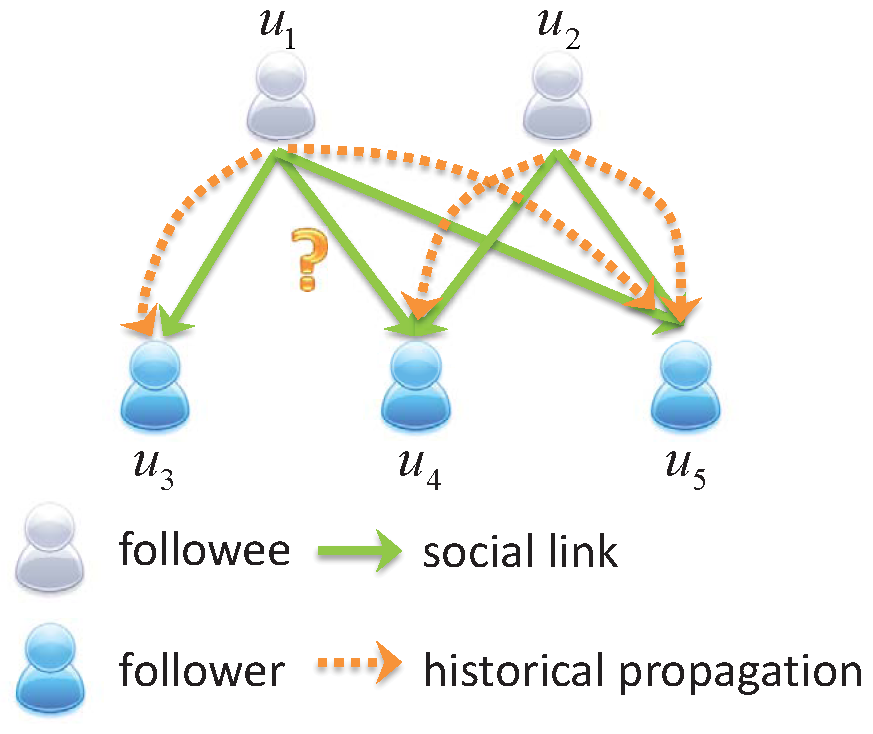}}
\subfigure[]{\includegraphics[width=0.275\textwidth]{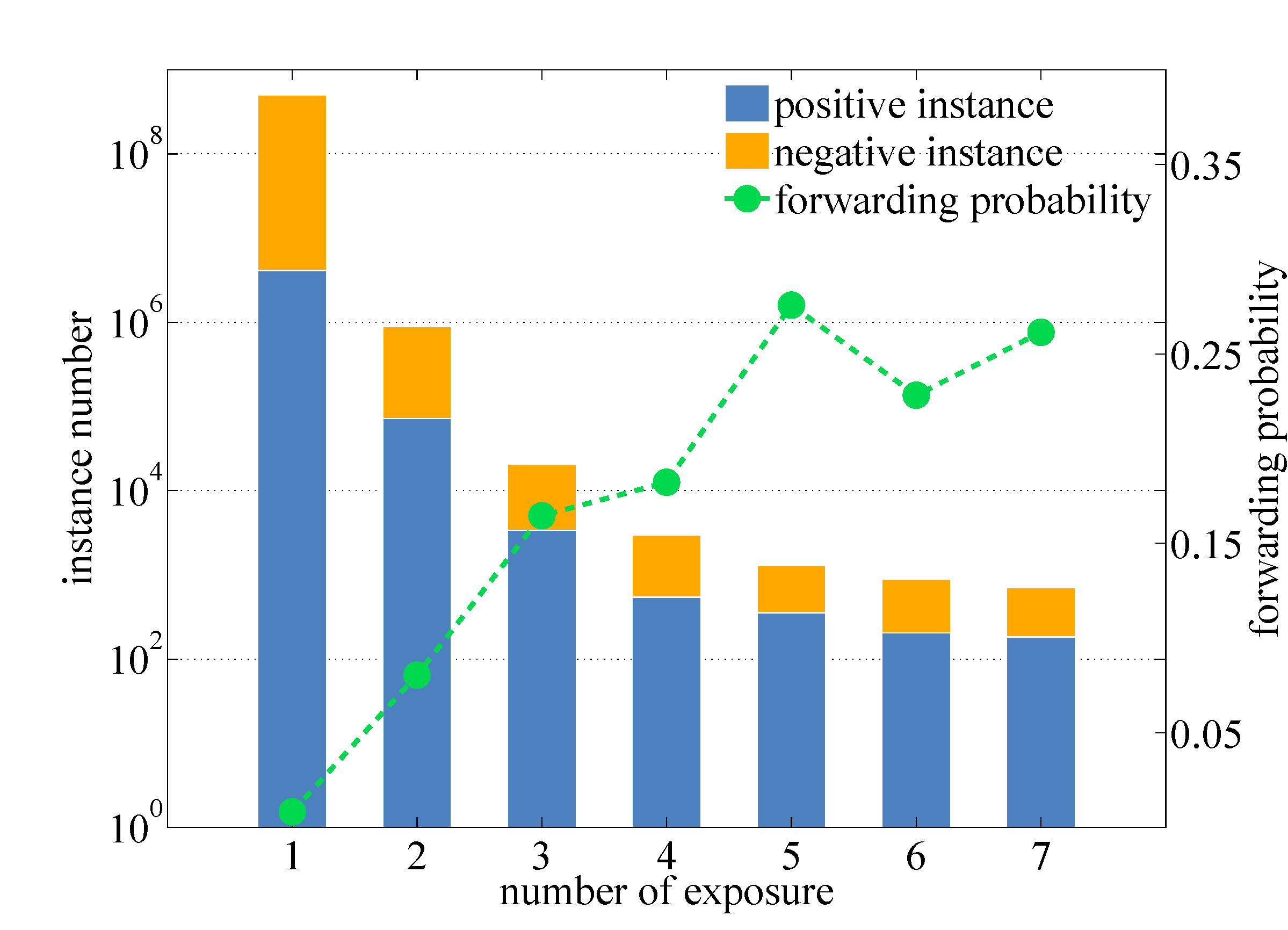}}
\caption{Motivations underlying our model. (a) Example of cascades to
illustrate the overfitting problem suffered by pair-wise models; (b)
Relationship between forwarding probability and number of exposures (observed
from Sina Weibo, Jan. 1-15, 2011). }
\label{fig:motivation}
\end{figure}

First, existing models suffer from severe overfitting problem, especially for
the pair of users without direct interactions. As shown in
Fig.~\ref{fig:motivation}(a), when messages are forwarded by users along social
links among them, not all social links matter in these message cascades. For
example, no forwarding behavior occurs between users $u_1$ and $u_4$, although
they have direct social link. In this case, existing models take the propagation
probability between them as zero, implying that it is never happened to
propagate information between the two users in the future. This overfitting
problem is actually caused by the hypothesis of existing models:
\emph{interpersonal influence between different pairs of users is independent
of each other}. This motivates us to adopt a user-specific manner for modeling
interpersonal influence among users, whose existence has been proved in
\cite{aral2012identifying,aral2009distinguishing}. Specifically, each user $u$
is modeled by two non-negative $d$-dimensional vectors: an influence vector $I_u$ and a
susceptibility vector $S_u$, characterizing the influence and susceptibility of
user $u$ over $d$ latent topics. For a pair of user $(u, v)$, the interpersonal
influence of $u$ on $v$ could be simply computed by the scalar product of $I_u$
and $S_v$. This concise representation requires only $2nd$ $(\ll n^2)$
parameters for $n$ users.

Second, the role of context in information propagation is rarely captured.
Existing models mostly assume that users are \emph{memoryless}, i.e., whether a
user forwards one message is not affected by her previous exposures to the
message. Indeed, this assumption is not supported by empirical observations
from real cascades of messages. Fig.~\ref{fig:motivation}(b) depicts the
relationship between the number of times $k$ that one user is exposed to a
message and the probability that the user will forward the message when she is
exposed to the message for the $k$-th time. As the number of exposures
increases, the forwarding probability increases accordingly: from 0.008 when
the number of exposures is 1 to 0.261 when the number of exposures is 5. This
observation clearly demonstrates that cumulative effect does exist in
information propagation, an effect also observed in many other
scenarios~\cite{baopeng2013cumulativeeffects,leskovec2007dynamics}. This
motivates us to adopt a context-dependent way to model the cumulative effect.

Given a message $m$, we denote its cascade $C^m$ with a chronological list of
activated users $(a_1^m,\ldots, a_N^m)$, where users are ranked in the ascending
order of the time they forward the message $m$. Whenever one user is activated (i.e.,
she posts or forwards the message), she has one chance to attempt to activate
other users. Whether her attempt succeeds depends on the \emph{cascade context}
at that time.

\begin{definition}{\textbf{Cascade context:}}
When one user $a_i^m$ ($i=1,\ldots,N$) becomes activated and she attempts to
activate a user $v$, the cascade context for this attempt is defined as
\begin{equation}
\mathcal{D}^m_{v,i}=\{a_j^m | j \leq i, \delta(a_j^m,v)=1\},
\end{equation}
where the indicator function $\delta(u, v)$ means whether $v$ could be
exposed to the message from $u$. In other words, cascade context means $v$'s
previous exposures to the message from other users. 
\end{definition}

In practice, it is difficult to exactly determine $\delta(u,v)$ since there is
no reliable mechanism to know whether user $v$ is exposed to a message forwarded
by user $u$, even when direct social link exists between them. In this paper, we
circumvent this problem using a delegate of the function $\delta$, i.e., the
aggregate diffusion network of historical cascades of messages. Detailed
discussions about diffusion network are given in experiments on real dataset.
Diffusion network characterizes the actual flow of information between users in
the past, providing us a good approximation for potential information
propagation in the future. More importantly, diffusion network is always
embedded in the collection of message cascades while social network is not
explicitly available in many scenarios. Therefore, the proposed model is
generally applicable to the scenarios with or without explicit social network.

With the above notations, we now model cascade dynamics of messages. For a
message $m$ with the cascade $(a_1^m, \ldots, a_N^m)$, each user could be
denoted by an $N$-dimensional status vector ${\bf z}_v^m$, with its element
$z^m_{v,j}$ indicating whether the user $v$ is in activated status right after
she is exposed to the message $m$ from user $a_j^m$. If user $v$ becomes activated when
she is exposed to the message $m$ from user $a_j^m$, we have $z^m_{v,i}=0$ $(1\leq i < j)$
and $z^m_{v,i}=1$ $(j \leq i \leq N)$. If $v$ keeps inactivated during the
cascade dynamics of message $m$, $z^m_{v,i}=0$ for any $i$.

\begin{figure}[t]
\centering
\includegraphics[width=0.42\textwidth]{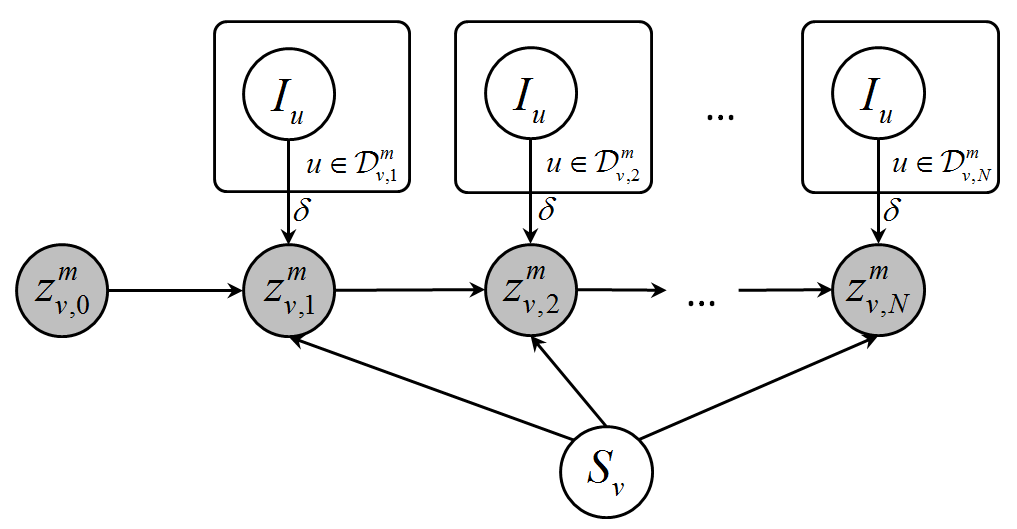}
\caption{Graphical representation of the proposed model.}
\label{fig:graphic_model}
\end{figure}

The likelihood of ${\bf z}^m_{v}$ is written as
\begin{equation}
P({\bf z}_v^m|\delta)=p\left( z^m_{v,0}\right) \prod\limits_{i=1}^N p\left(
z^m_{v,i} | z^m_{v,i-1}, \mathcal{D}^m_{v,i}, \delta \right),
\label{eq:user_status}
\end{equation}
where $z^m_{v,0}$ is introduced only for simplifying the notation. Here the
first term depicts whether $v$ is the source of message $m$, defined as
\begin{equation}
\label{eq:source_message}
p\left( z^m_{v,0}=1 \right) = \left\{
\begin{aligned}
&1, v \text{ is the source}\\
&0, \text{otherwise}
\end{aligned} 
\right. .
\end{equation}
The second term characterizes the transition of user's status, formally defined
as
\begin{equation}
\label{eq:prop_prob}
\small
\begin{aligned}
p\bigl( z^m_{v,i}=1 |& z^m_{v,i-1}=1, \mathcal{D}^m_{v,i}, \delta \bigr) = 1, \\
p\bigl( z^m_{v,i}=1 |& z^m_{v,i-1}=0, \mathcal{D}^m_{v,i}, \delta \bigr) = \\
&1-\exp\bigl({-\lambda \delta\bigl( a_i^m,v \bigr)
\sum\nolimits_{u\in \mathcal{D}^m_{v,i}} I_u^T S_v}\bigr), \\
p\bigl( z^m_{v,i}=0 |& z^m_{v,i-1}=0, \mathcal{D}^m_{v,i}, \delta \bigr) =  \\
&1-p\bigl( z^m_{v,i}=1 | z^m_{v,i-1}=0, \mathcal{D}^m_{v,i}, \delta \bigr),
\end{aligned}
\end{equation}
where $\lambda$ is a scaling factor, modulating the effect of cascade
context. The first equation implies that a user cannot become inactivated as
long as she is activated. The other two equations model how the transition
probability is influenced by cascade context, reflected by a cumulative manner of interpersonal
influence. The factor $\delta(a^m_i,v)$ determines whether the user $v$ could be
exposed to the message $m$ from user $a^m_i$. For clarity, we give a graphical
model representation in Fig.~\ref{fig:graphic_model} to illustrate the process that
user $v$'s status changes with the dynamics of message $m$.

Assuming independent cascades, the likelihood of all cascades ${\bf C}$ is
a product of the likelihoods given by Eq.~(\ref{eq:user_status})
\begin{equation}
L\left( {\bf C} \right)=\prod\limits_{m=1}^{|{\bf C}|} \prod\limits_{v\in V}
P\left( {\bf z}_v^m | \delta \right).
\end{equation}

The parameters of the model are learned by minimizing the negative
logarithmic likelihood, namely loss function, of all the collection of cascades as
the following objective function
\begin{equation}
\label{eq:obj_func}
\mathcal{L}({\bf C}) = -\sum_{m=1}^{|{\bf C}|}{\sum_{v\in V} \sum_{i=1}^N{\log
p\left( z^m_{v,i} | z^m_{v,i-1}, \mathcal{D}^m_{v,i}, \delta \right)}}.
\end{equation}
Note that $p\left( z^m_{v,0} \right)$ is always 1, hence omitted in
Eq.~(\ref{eq:obj_func}).

\section{Parameter estimation}
\label{sec:solution}
In this section, we develop algorithm to estimate the parameters of
the proposed model. Generally speaking, parameter estimation could be completed
via directly minimizing Eq.~(\ref{eq:obj_func}) with respect to $I$ and $S$.
However, the huge number of possible configurations of cascade contexts in
Eq.~(\ref{eq:obj_func}) results in high computational cost. Indeed,
\emph{one} cascade context could repeatedly occur in many cascades, causing huge
duplicated computation of Eq.~(\ref{eq:prop_prob}).

Here we propose to reduce the duplicated computation by taking advantage of
overlapped cascade contexts among multiple cascades. Firstly, we introduce some
symbols to simplify the optimization. Let $\mathcal{P}(v)$ be the set of
possible configurations of cascade contexts pertaining to $v$, occurred in all
cascades. For example, as shown in Fig.~\ref{fig:motivation}(a), $\{u_1\}$,
$\{u_2\}$ and $\{u_1,u_2\}$ are all the cascade contexts for user $u_5$. We
group cascade contexts in terms of users and re-organize the logarithmic
likelihood in Eq.~(\ref{eq:obj_func}) as
\begin{equation}
\label{eq:re_obj_func}
\resizebox{.43\textwidth}{!}{
$
\mathcal{L}({\bf C})\!=\!-\!\!\!\sum\limits_{v\in
V}\!\sum\limits_{\mathcal{D}_{v,i}\in \atop \mathcal{P}(v)}\!\!\!\! \left({
n_{z_{v,i},{\mathcal{D}_{v,i}}}\log p\left( z_{v,i} |
z_{v,i-1},\mathcal{D}_{v,i},\delta \right)} \right), $
}
\end{equation}
where $\mathcal{D}_{v,i}$ refers to one configuration of cascade context for
user $v$, independent of specific cascade, and $n_{z_{v,i},{\mathcal{D}_{v,i}}}$ counts 
the frequency of the configuration $\mathcal{D}_{v,i}$ emerged in all cascades,
relative to user $v$'s status $z_{v,i}$.

Parameters estimated by directly minimizing the logarithmic likelihood in
Eq.~(\ref{eq:re_obj_func}) may suffer from overfitting problem, a common
problem in likelihood maximization estimation. To combat this problem, we
regularize the parameter vectors $I$ and $S$ and obtain the final objective
function for parameter estimation
\begin{equation} 
\resizebox{.43\textwidth}{!}{
$
\begin{aligned}
\mathcal{\hat{L}}({\bf C})\!=\!& -\!\!\sum\limits_{v\in
V}\!\sum\limits_{\mathcal{D}_{v,i}\in \atop \mathcal{P}(v)}\!\!\!\! \left({
n_{z_{v,i},{\mathcal{D}_{v,i}}} \log p\left( z_{v,i} |
z_{v,i-1},\mathcal{D}_{v,i},\delta \right)} \right) \\
& + \gamma_{I}\|I\|^2_F + \gamma_{S}\|S\|^2_F , \\
\textit{s.t. } & I_{ij}  \geq~0, S_{ij} \geq 0, \forall {i,j},
\end{aligned}
$
}
\end{equation}
where $\gamma_I$ and $\gamma_S$ are regularization
parameters, and ${\|\cdot\|}_F$ is Frobenius norm.

\begin{algorithm}[t]
\caption{Parameter estimation}
\small
\label{alg:nmf}
\begin{algorithmic}
\STATE {\bf Input:} Collection of cascades observed in a given time period, the
maximum epoch $M$, and regularization parameters $\gamma_I$ and $\gamma_S$
\STATE {\bf Output:} User-specific influence and susceptibility $I$, $S$
\STATE
\STATE Construct diffusion network $\delta$ from cascades 
\STATE Initialize parameters with random values, including $I$, $S$
\REPEAT
	\FOR{$i = 1$ to $n$}
		\STATE Calculate gradients ${{\partial \mathcal{\hat{L}}}
		 \mathord{\left/ {\vphantom {{\partial \mathcal{\hat{L}}} {\partial \mathcal{\hat{L}}}}} \right.
		 \kern-\nulldelimiterspace} {\partial I_u}}$ and ${{\partial
		 \mathcal{\hat{L}}} \mathord{\left/ {\vphantom {{\partial \mathcal{\hat{L}}} {\partial \mathcal{\hat{L}}}}} \right.
		 \kern-\nulldelimiterspace} {\partial S_v}}$
	\ENDFOR
	\STATE Update $I$ and $S$ with PG method
\UNTIL maximum epoch $M$ is reached or gradients vanish
\end{algorithmic}
\end{algorithm}

Finally, using \textit{Projected Gradient} (PG)
method~\cite{lin2007projectgradient}, we develop an iterative algorithm 
for parameter estimation, leveraging the gradients with respect to $I$ and
$S$ as
\begin{equation}
\resizebox{.43\textwidth}{!}{
$
\begin{aligned}
\frac{\partial \mathcal{\hat{L}}}{\partial I_u} 
=&{-}\lambda\sum_{v\in V}\!S_v
\sum_{\mathcal{D}_{v,i}\in
\atop \mathcal{P}(v)} \! \mathcal{I}_{u\in \mathcal{D}_{v,i}}
\Bigl( n_{z_{v,i}=1,\mathcal{D}_{v,i}}
\frac{1-p_{v,{\mathcal{D}_{v,i}}}}{p_{v,{\mathcal{D}_{v,i}}}} \\	
&-n_{z_{v,i}=0,\mathcal{D}_i(v)} \Bigr)+\gamma_I I_u,\\
\frac{\partial \mathcal{\hat{L}}}{\partial S_v}
=&{-}\lambda\sum\limits_{\mathcal{D}_{v,i}\in
\atop \mathcal{P}(v)} \! \sum\limits_{u\in \mathcal{D}_{v,i}}
\!I_{u} \Bigl(n_{z_{v,u}=1, \mathcal{D}_{v,u}}
\frac{1-p_{v,{\mathcal{D}_{v,u}}}}{p_{v,\mathcal{D}_{v,u}}} \\
&-n_{z_{v,u}=0,\mathcal{D}_{v,u}} \Bigr)+\gamma_S S_v,
\end{aligned}
$
}
\end{equation}
where $\mathcal{I}$ is an indicator function, and $p_{v,{\mathcal{D}_{v,i}}}$
is a concise form of $p\left( z_{v,i}=1 |
z_{v,i-1},\mathcal{D}_{v,i},\delta \right)$.
The algorithm for parameter estimation is described in Algorithm~\ref{alg:nmf}.

\section{Experiments}
\label{sec:experiments}
We evaluate LIS model on both synthetic data and real world data, i.e.
microblogging data from Sina Weibo. Here we only choose scalable modeling
methods as baselines, i.e., expectation maximization estimation
(EM)~\cite{saito08}, static Bernoulli model (SB), and static Jaccard model
(SJ)~\cite{goyal2010learning}. To reduce the overfitting problem suffered from
these models, we apply \textit{matrix factorization} (MF)
method~\cite{salakhutdinov2008probabilistic} as a post-processing improvement,
forming a stronger baseline. We demonstrate the benefit of the LIS model at
predicting the cascade dynamics, cascade size, and ``who will be retweeted''.
Finally, we analyze the topics associated with the learned user-specific latent
influence and susceptibility.

\subsection{Experiments on synthetic data}
To validate whether the proposed algorithm could obtain good estimation of
user-specific influence and susceptibility, we first conduct tests on synthetic
data, where cascades are generated according to parameters known a prior. These
tests also offer us some intuitions about difficulties at predicting cascade dynamics.

\subsubsection{Experimental setup.}
We first generate two synthetic diffusion networks: one is constructed using
Barab\'asi-Albert (BA) model~\cite{barabasi1999emergence}, denoted as original
network; the other is generated by shuffling the original
network~\cite{Molloy95}, denoted as shuffle network. We set
the dimensionality of user's influence $I_u$ and susceptibility $S_u$ as 5,
and sample $I_u$ and $S_u$ from $f({\bf{x}})=1/2\sqrt{{\bf{x}}}, {\bf{x}}\sim
U(0,1)^5$, where $U(0,1)$ refers to a uniform distribution. We then generate
cascades over the two networks according to our LIS model with $\lambda=0.01$.
We take 80\% of the cascades generated from original network as training
dataset, and the rest of cascades generated from the original network and
the cascades from the shuffle network as test dataset. The two test datasets
have equal size, offering a fair comparison.

\subsubsection{Predicting cascade dynamics.}
We deal with this problem as a set of binary classification
problem, predicting whether one user will be activated
under specific cascade context. Thus we use AUC (the area of under the ROC
curve) as the evaluation metric~\cite{Fawcett2006}. As a reference for the performance
comparison, we offer an upper bound of our LIS model (denoted as UB): predicting
cascade dynamics according to the parameters $I$ and $S$ that are used to
generate cascades in test data.

Results are listed in Table~\ref{tb:toy_auc_curve}. It is seen that LIS
model consistently outperforms baselines, and its AUC value is very close to UB,
when predicting on original network. Particularly, the performance of LIS model
is stable on both original and shuffle networks, while baseline methods suffer
from much performance reduction (some even close to 0.5, equal to random guess
approach), resulted from the overfitting problem of pair-wise models.

\begin{table}[t]
\centering
\caption{Cascade dynamics prediction on synthetic data}
\label{tb:toy_auc_curve}
\begin{tabular*}{0.435\textwidth}{c|ccccc} \hline
network & UB & LIS & SB & SJ & EM \\ \hline
original & 0.659 & 0.654 & 0.607 & 0.618 & 0.561 \\ 
shuffle & 0.659 & 0.608 & 0.509 & 0.525 & 0.507 \\
\hline
\end{tabular*}
\end{table}

\subsection{Experiments on Microblog}
\subsubsection{Dataset.}
\label{sec:dataset}
The Microblog data from Sina Weibo website is published by WISE 2012
Challenge\footnote{http://www.wise2012.cs.ucy.ac.cy/challenge.html}, spanning
from January 1, 2011 to Feburary 15, 2011.
We extract the cascade records posted between January 1, 2011 and February
15, 2011, and split the extracted records
into three training datasets, i.e., D1, D2, D3, each persisting a period of
half a month. Furthermore, for each training dataset, we extract the cascade
records in the following 5 days as test, i.e., T1, T2 and T3. We only consider the users
who appear in all the three training datasets, obtaining
199,408 users. Dataset statistics are depicted in
Table~\ref{tb:data_desc}. We conduct an empirical study to demonstrate the
severity of overfitting problem. 
Over 70\% of \textit{forwarding traces}---historical
paths for information flows---in test data are never observed in
training data, posing a big challenge to previous pair-wise models, when applied in real data.

\begin{table}[t]
\centering
\caption{Dataset statistics}
\label{tb:data_desc}
\resizebox{0.47\textwidth}{!}{
\begin{tabular}{c|cc|c|cc} \hline
& \multicolumn{2}{c|}{training data} & & \multicolumn{2}{c}{test data} \\
 & cascades & period &  & cascades & period \\
\hline
D1 & 395,852 & 01/01-01/15 & T1 & 160,868 & 01/16-01/20 \\
D2 & 453,356 & 01/16-01/31 & T2 & 122,509 & 02/01-02/05 \\
D3 & 386,152 & 02/01-02/15 & T3 & 145,143 & 02/16-02/20 \\ \hline
\end{tabular}}
\end{table}

\subsubsection{Diffusion network.}
\begin{figure}[t]
\centering
\includegraphics[width=0.47\textwidth]{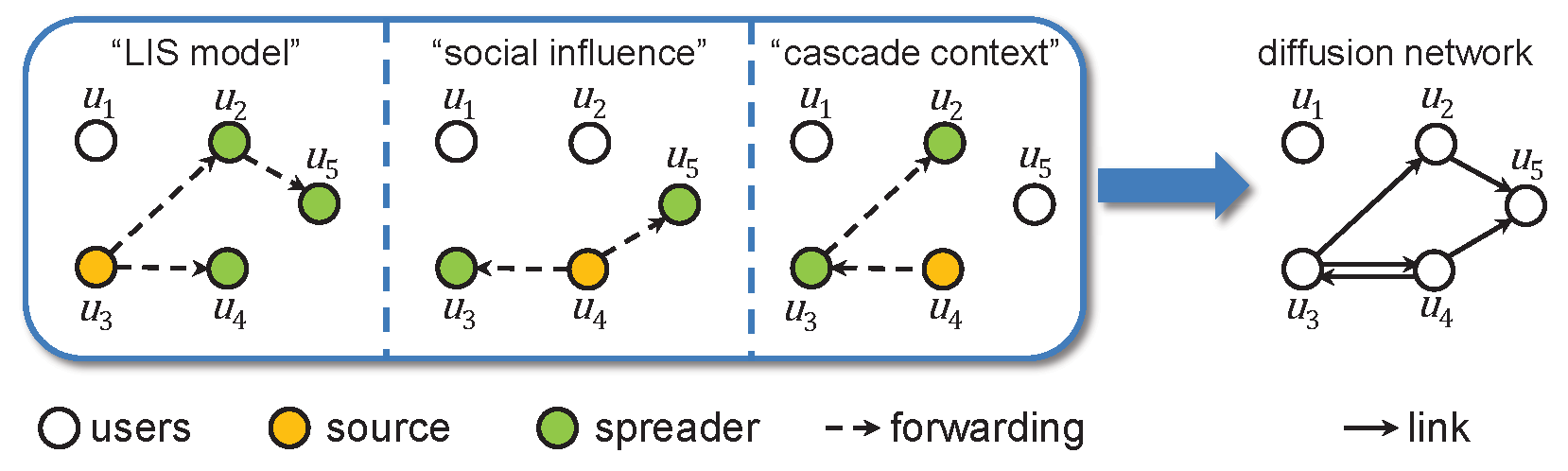}
\caption{Illustration of constructing diffusion network. The
left box contains three observed cascades, and the right is their
aggregated diffusion network.}
\label{fig:cascade_explain}
\end{figure}

Exact diffusion network is hard to obtain, since there is no clear
clues indicating whether a user is exposed to a message forwarded by her
followee. Previous works focus on directly inferring diffusion
networks, which is not applicable to large-scale scenarios~\cite{gomez2010inferring,gomez2013modeling,du2012learning,KurashimaKDD2014}.
Here we estimate diffusion network according to a large collection of historical
cascades: one cascade has a	 collection of forwarding traces over the period of
observation, forming a directed graph. We aggregate these graphs of all cascades into a
diffusion network, providing us a good approximation for potential information propagation
in the future. Figure~\ref{fig:cascade_explain} gives an example to illustrate
the construction of diffusion network.

\begin{figure*}[t]
\centering
\subfigure[T1]{\includegraphics[width=0.3\textwidth]{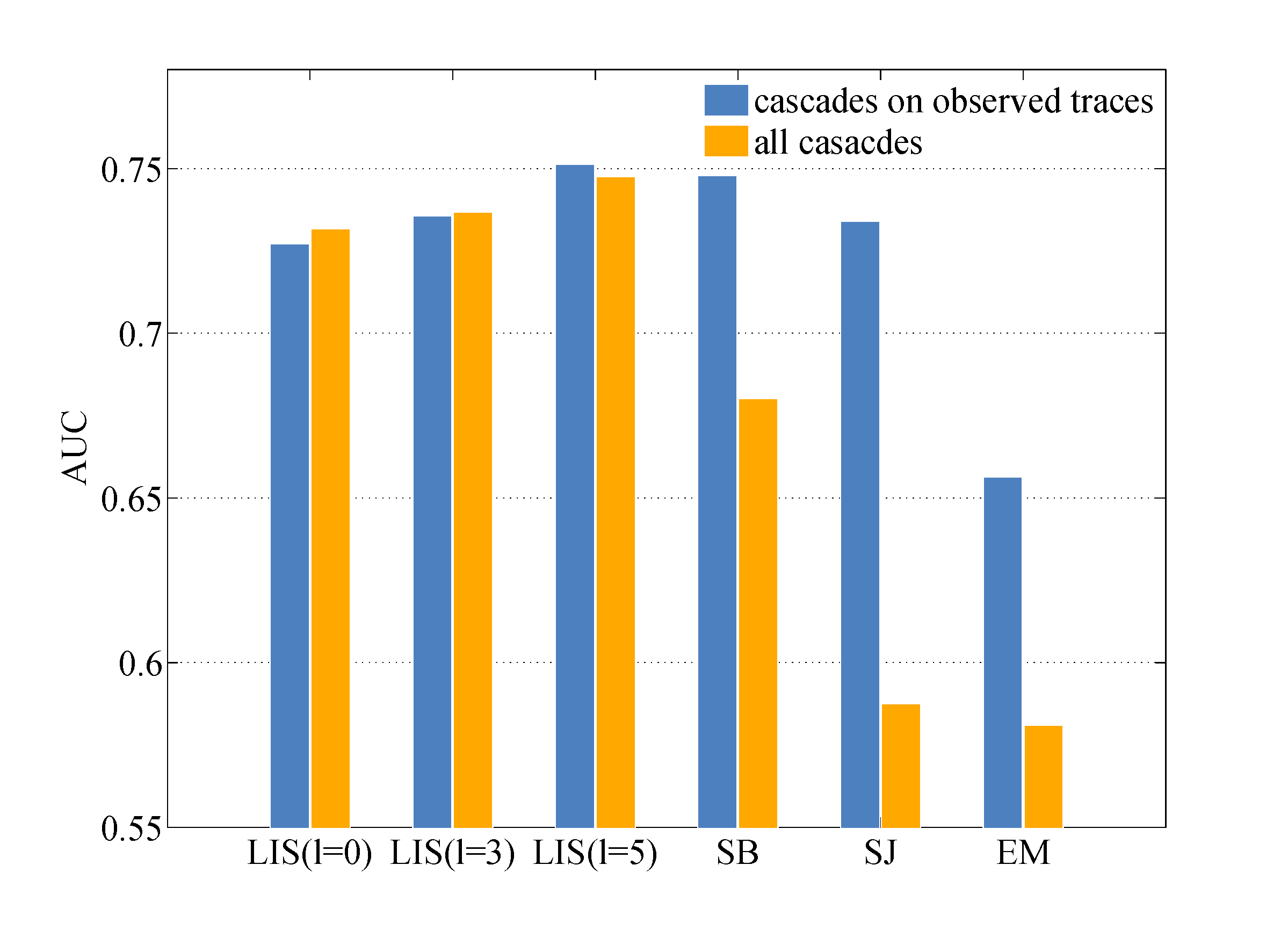}}
\subfigure[T2]{\includegraphics[width=0.3\textwidth]{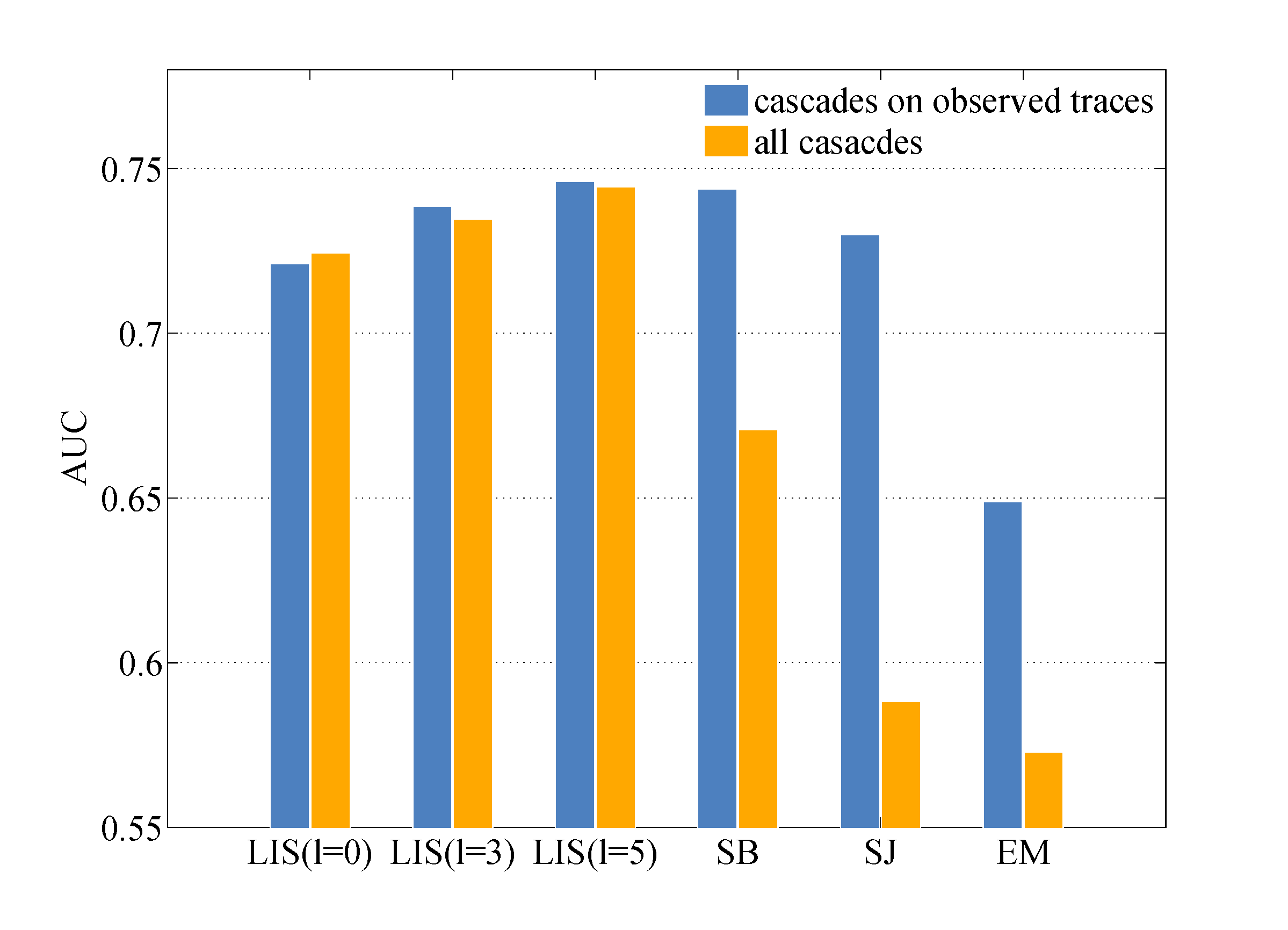}}
\subfigure[T3]{\includegraphics[width=0.3\textwidth]{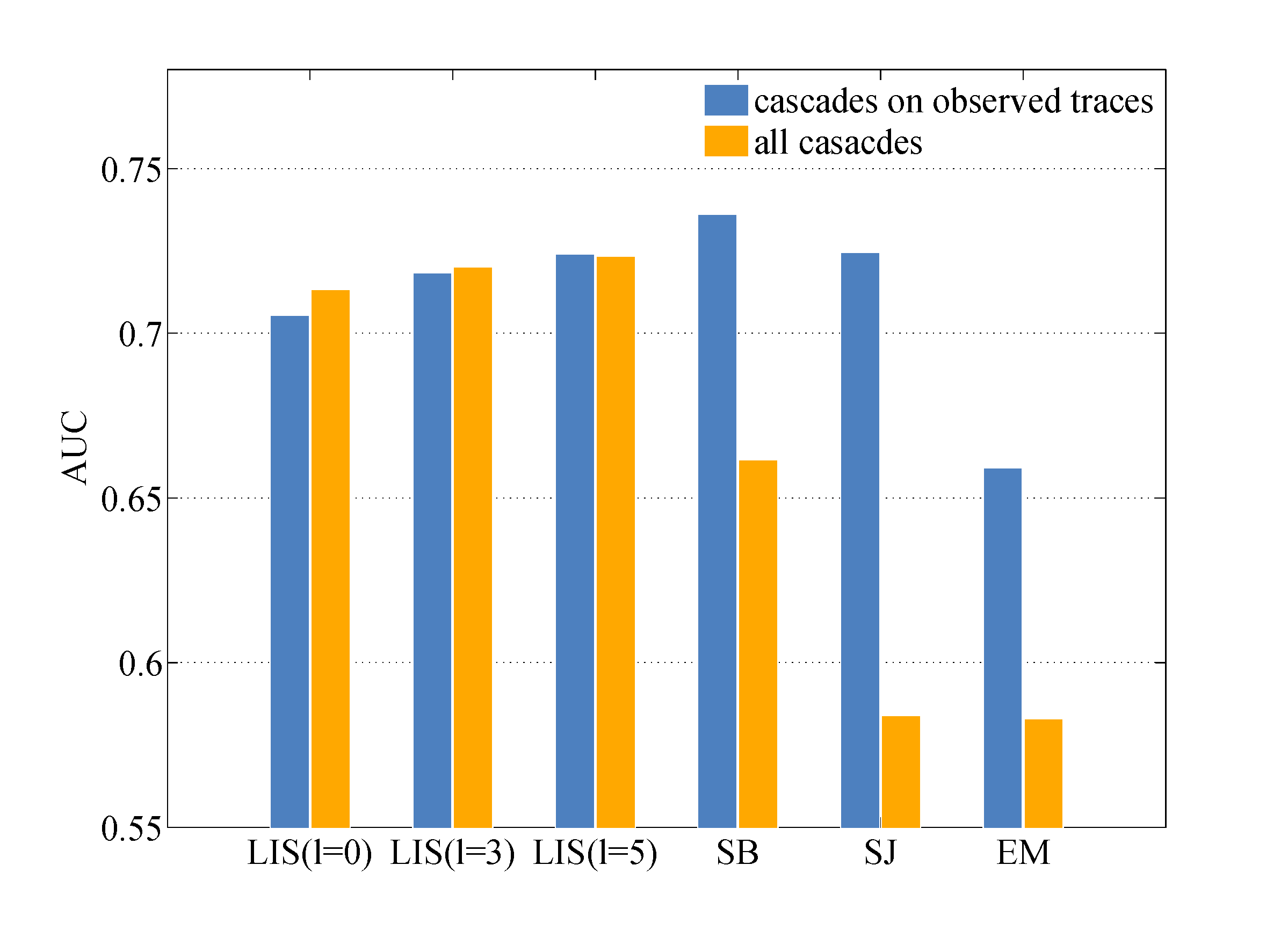}}
\caption{AUCs of cascade dynamics prediction on microblog.}
\label{fig:AUC_DCP}
\end{figure*}

\subsubsection{Predicting cascade dynamics.}
We introduce the length of cascade context $l$ to capture user's
context-dependent level for exploring context-dependent factors like cumulative
effects in information propagation. When a user decides whether or not to
forward a message, only the recent $l$ exposures take effects in cascade
context.
The case $l=0$ implies that users are memoryless, i.e., without cumulative effect,
reproducing the independent assumption~\cite{kempe2003maximizing}. To illustrate the
difference of prediction performance at combating overfitting problem, we
additionally evaluate the prediction only over observed forwarding traces. The
experimental results are presented in Fig.~\ref{fig:AUC_DCP}. Despite that
baselines can better handle prediction instances over observed forwarding
traces, the AUCs decrease dramatically when applied to the prediction over all
forwarding traces in test data. It means that pair-wise models, even with an
improvement using matrix factorization, still suffer from overfitting problem when a large
proportion of forwarding traces in test data are unobserved.
LIS model shows stable performance, indicating that it works consistently better than baseline models,
no matter whether the forwarding traces are happened in history. Furthermore, we
explore the effects of different length of cascade context $l$ in LIS model.
With different settings of $l$, LIS model performs better as the increase of
$l$, shown in both two types of AUCs. It agrees with the empirical observation
on cumulative effect of information
propagation~\cite{ugander2012structural,baopeng2013cumulativeeffects,leskovec2007dynamics},
and LIS model can effectively capture the effects.

\begin{table*}[t]
\centering
\caption{MAPEs of cascade size prediction}
\label{tb:cas_size}
\begin{tabular*}{0.87\textwidth}{ccccccc} \hline
& LIS ($l=0$) & LIS ($l=3$) & LIS ($l=5$) & SB & SJ & EM\\ \hline
T1 & 0.163$\pm$0.0133 & 0.140$\pm$0.0155 & 0.141$\pm$0.0217 &
0.191$\pm$0.0190 & 0.524$\pm$0.0046 & 0.258$\pm$0.0160 \\
T2 & 0.287$\pm$0.0093 & 0.280$\pm$0.0080 & 0.286$\pm$0.0065 &
0.333$\pm$0.0099 & 0.621$\pm$0.0048 & 0.338$\pm$0.0387
\\
T3 & 0.095$\pm$0.0150 & 0.094$\pm$0.0150 & 0.097$\pm$0.0093 & 0.171$\pm$0.0388
& 0.505$\pm$0.0450 & 0.189$\pm$0.0112 \\
\hline
\end{tabular*}
\end{table*}

\subsubsection{Cascade size prediction.}
Cascade size prediction, as a key part of influence maximization and viral
marketing, is one of the most important applications based on modeling
cascade dynamics~\cite{bao2013popularity,ShenAAAI14}. To guarantee that all
cascades propagate sufficiently from their message sources, we only choose such cascades
that are initially posted at the first day in each test data for cascade
size prediction. Starting from the true sources of cascades in training dataset,
we simulate cascade dynamics using the learned
LIS model as a prediction. We group simulated cascades into
bins according to their cascade size. By counting the number
of cascades in each bin, we get a vector of counts indexed by
cascade sizes. In the same way, we also get the vector of counts from
the real cascades in test data as the ground truth, and we smooth out those size
bins containing few cascades for statistical significance. Thus
the cascade size prediction can be evaluated by \textit{mean absolute
percentage error} (MAPE), where a smaller value indicates a better prediction.
Since the cascades are generated by simulations, we repeat the prediction,
recording the average values and standard deviations of MAPE in
Table~\ref{tb:cas_size}. It is seen that all LIS models with different settings
of $l$ achieve better MAPE values than baselines on test data, indicating LIS
model can estimate completed cascades propagation more efficiently than
pair-wise models. Particularly, the MAPEs in test data T2 are
much larger than the values in the other two test data. One possible explanation
is that these cascades in test data T2 span Chinese New Year (from
Jan. 2 to Jan. 8, 2011) when users were dominated by offline social activities.
Thus, it is hardly to observe completed cascades in the short
period, resulting in unexpected performance by simulating complete cascades
without accounting these factors.

\subsubsection{Prediction of ``who will be retweeted''.}
The problem ``who will be retweeted'' is a way to examine interpersonal
influence under quantitative understanding. In the scene of multi-exposures,
high interpersonal influence will have high probability to be
forwarded. LIS model provides a direct quantitative metric for
interpersonal influence by the scalar product of $I_i$ and
$S_j$ between user $i$ and $j$. Propagation probability is another metric of
interpersonal influence for traditional pair-wise models. We therefore
deal with the prediction task as a ranking problem of interpersonal
influence. The user with higher rank is more probable to be 
retweeted. We evaluate the prediction performance by metrics of average
\textit{Accuracy} (Acc) of top-1 prediction and \textit{Mean Reciprocal Rank}
(MRR)~\cite{voorhees1999trec}. The larger values of Acc and MRR indicate better
predictions. The results are given in Table~\ref{tb:acc_retweet_whom}. It is
seen that LIS model achieves a much better Acc and MRR than the baselines, which
indicates that LIS model provides a more efficient perspective to inspect
interpersonal influence in information cascades.

\begin{table}[t]
\centering
\caption{Accuracies and MRRs of prediction of ``who will be retweeted''}
\label{tb:acc_retweet_whom} 
\begin{tabular}{cc|cccc} \hline
& & LIS ($l=5$) & SB & SJ & EM \\ \hline
\multirow{3}{*}{{\textit{Acc($\%$)}}} & T1 & 58.48 & 57.02 &
49.99 & 53.48
\\
& T2 & 57.61 & 55.05 & 49.65 & 52.23 \\
& T3  & 59.58 & 55.38 & 50.85 & 55.41 \\ \hline
\multirow{3}{*}{{\textit{MRR}}} & T1 & 0.791 & 0.784 & 0.748 &
0.766
\\
& T2 & 0.786 & 0.773 & 0.745 & 0.758 \\ 
& T3 & 0.797 & 0.775 & 0.752 & 0.775 \\ \hline
\end{tabular}
\end{table}

\subsubsection{Topic allocation in latent features} 
\begin{figure}[t]
\centering
	  \subfigure[influence]{\includegraphics[width=0.23\textwidth]{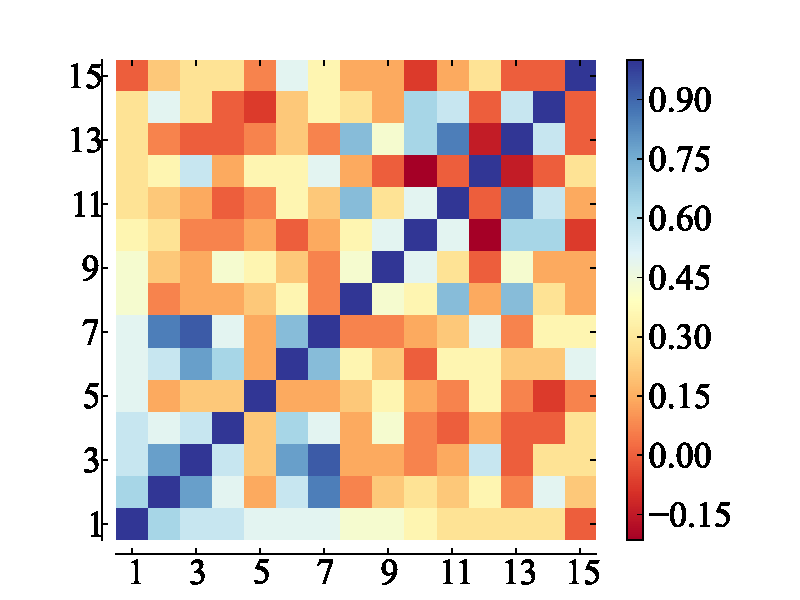}}
	  \subfigure[susceptibility]{\includegraphics[width=0.23\textwidth]{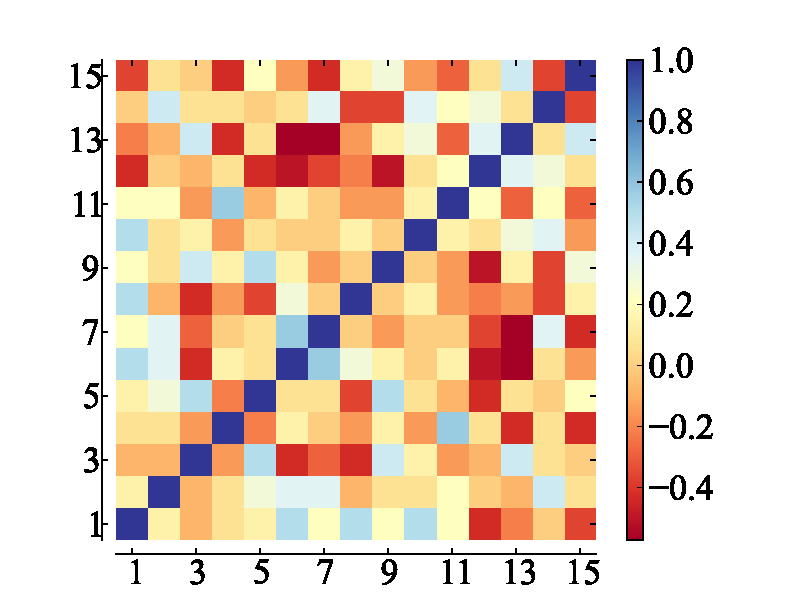}}
	  \caption{Kendall-$\tau$ rank correlation coefficient.}\label{fig:feat_dif}
\end{figure}

To explore topic allocation in LIS model, we select top 10,000 users
for each dimension of latent influence and susceptibility features, according to
their influence and susceptibility respectively.
We use hashtags as the topic of these messages posted by the selected users.
Messages without hashtags are discarded. We keep 8 hashtags with statistical
significance in terms of message number. Then we calculate the distribution of
each remained hashtags in latent features according to the frequency used by users.
Next, we rank hashtags by the results of distribution in latent features
and utilize kendall-$\tau$ rank correlation coefficient~\cite{kendall1938new}
to analyze the difference between pairs of features. Figure~\ref{fig:feat_dif}
shows the heat map of difference between pairs of features. Every grid
represents the correlation between two features. The coefficient values close to
1 or -1 refer to positive and negative correlation between two features, depicted in
deep blue or deep red respectively. If two features are
independent, the coefficient value approaches to 0, colored by yellow. As shown
in Fig.~\ref{fig:feat_dif}, most pairs of features are approximatively
independent in both influence and susceptibility, indicating that the learned
features are highly discriminative in topic level. Furthermore, we illustrate
the top 3 hashtags in latent features shown in Table~\ref{tb:feat_exm},
containing four pairs of features and their hashtags. The influence features 3
and 7 are coherent well with high kendall-$\tau$ coefficients 0.93, as hashtag
``Xiaomi release'' is on the top of both. Those influence and susceptibility
features with lower or zero kendall-$\tau$ coefficients have distinguished
hashtags.

\begin{table}[th!]
\caption{Topic allocation in features}
\label{tb:feat_exm}
\centering
\subtable[two pairs of features in influence]{
\resizebox{0.47\textwidth}{!}{
\begin{tabular}{ccl} 
\hline
kendall-$\tau$ & \multirow{2}{*}{feature index} & \multirow{2}{*}{ranked hash
tags} \\
coefficients & & \\ \hline
\multirow{4}{*}{0.93} & \multirow{2}{*}{3} & 1. Xiaomi release; 2. Tang Jun
education \\
& & qualification fake; 3. House prices \\
& \multirow{2}{*}{7} & 1. Xiaomi release; 2. Qian Yunhui; \\
& & 3. House prices \\ \hline
\multirow{4}{*}{0} & \multirow{2}{*}{2} & 1. Yao Ming retire; 2. Case of
running fast \\
& & car in Heibei University; 3. Xiaomi release \\
& \multirow{2}{*}{14} & 1. Xiaomi release; 2. Qian Yunhui; \\
& & 3. House prices \\ \hline
\end{tabular}
}
}
\qquad
\subtable[two pairs of features in susceptibility]{
\resizebox{0.47\textwidth}{!}{
\begin{tabular}{ccl} 
\hline
kendall-$\tau$ & \multirow{2}{*}{feature index} & \multirow{2}{*}{ranked hash
tags} \\
coefficients & & \\ \hline
\multirow{5}{*}{0.43} & \multirow{3}{*}{6} & 1. Incident of self-burning at
Yancheng, \\
& & Jiangsu; 2. Tang Jun education qualifica-\\
& & tion fake; 3. Yao Ming retire \\
& \multirow{2}{*}{11} & 1. Xiaomi release; 2. Qian Yunhui; \\
& & 3. House prices \\ \hline
\multirow{6}{*}{0} & \multirow{4}{*}{4} & 1. Incident of self-burning at
Yancheng, \\
& & Jiangsu; 2. Tang Jun education qualifica- \\
& & tion fake; 3. Case of running fast car in \\
& & Heibei University \\
& \multirow{2}{*}{7} & 1. Qian Yunhui; 2. Incident of self-burn-\\
& & ing at Yancheng, Jiangsu; 3. House prices \\ \hline
\end{tabular}
}
}
\end{table}

\section{Conclusions}
\label{sec:conclusions}
In this paper, we proposed a concise probabilistic model for the information
propagation on social network, explicitly characterizing the
influence and susceptibility of each user with two low-dimensional vectors
respectively. The proposed model distinguishes itself from previous models at
its capability of modeling both the interpersonal influence between any pair of
users and the cumulative effect in information propagation. We also designed
effective algorithms to train the model based on maximizing logarithmic
likelihood of information cascades. Our model does not require the knowledge of
social network structure, hence having wide applicability to the scenarios with
or without explicit social networks. We evaluated the effectiveness of our model
on synthetic dataset and a large-scale microblogging dataset from Sina Weibo,
the largest social media in China. Experimental results demonstrate that our
model consistently outperforms existing methods at predicting
cascade dynamics, cascade size, and ``who will be retweeted''. Moreover, the
learned user-specific influence and susceptibility vectors provide us a
quantitative way to understand topic-related interpersonal influence in information propagation.


\section{Acknowledgments}
This work was funded by the National Basic Research Program of China under
grant number 2012CB316303, the National High-tech R\&D Program of China under
grant number 2014AA015103, and the National Natural Science Foundation of China
with Nos 61232010, 61472400, 61202213, 61272536. This work is also partly
funded by the Beijing Natural Scientific Foundation of China under grant number 4122077.
The authors thank to the members of NASC research group
(\url{www.groupnasc.org}) for valuable discussions and suggestions.



\bibliographystyle{aaai}
\bibliography{social}

\end{document}